\documentclass[10pt]{article}

\usepackage{amsmath}
\usepackage{amssymb}
\usepackage{graphics}
\usepackage{rotating}
\usepackage{cite}
\usepackage{color}
\usepackage{fancybox}

\usepackage{setspace}

\def\0{\mbox{\tiny $0$}}
\def\1{\mbox{\tiny $1$}}
\def\2{\mbox{\tiny $2$}}
\def\3{\mbox{\tiny $3$}}
\def\4{\mbox{\tiny $4$}}
\def\5{\mbox{\tiny $5$}}
\def\6{\mbox{\tiny $6$}}
\def\7{\mbox{\tiny $7$}}
\def\8{\mbox{\tiny $8$}}
\def\9{\mbox{\tiny $9$}}
\def\I{\mbox{\tiny $I$}}
\def\II{\mbox{\tiny $II$}}
\def\III{\mbox{\tiny $III$}}
\title{\shadowbox{\large \bf RELATIVE HELICITY PHASES IN PLANAR DIRAC SCATTERING}}

\author{
\small  Stefano De Leo\thanks{Department of Applied Mathematics,
State University of Campinas, Brazil [deleo@ime.unicamp.br] } \,\,
and\, Pietro Rotelli\thanks{Department of Physics, University of
Salento and INFN Lecce, Italy [rotelli@le.infn.it]}}

\date{\small
\fcolorbox{black}{yellow} {\color{red} $\bullet$ {\color{black}{
{\footnotesize  {\sc Physical Review A} {\bf 86}, 032113-5 (2012)}}}
{\color{red}{$\bullet$}} } }

\begin{document}
%


\maketitle

\vspace*{-.7cm}

\begin{abstract}
\noindent We study planar Dirac scattering for an electrostatic  stratified
barrier potential. The general expressions for
transmitted and reflected waves are derived. Of particular
interest is the information upon relative helicity phases. We also briefly discuss some possible applications.
\end{abstract}












\section*{\normalsize I. INTRODUCTION}

One of the  objectives of this work is the derivation of the
general formulas for spinor plane wave scattering from a stratified
barrier potential. The potential is
assumed to be the zero component of a four vector potential, also
known in the literature as an ``electrostatic" potential \cite{R01}. Our
approach is three dimensional but the assumed stratified direction ($x$ axis)
of the potential together with the incoming three momentum
determines the scattering plane within which lie both the
reflected and transmitted momenta. We shall call this the $x$-$y$
plane and hence, without lose of generality, we consider planar
scattering, both diffusion and tunnelling \cite{R0301,R0302,R0303,R0304,R0305,R0306,R0307,R0308,R0309,R0310,R0311,R0312,R0313}. Our spinors are Dirac
spinors. We shall also operate with helicity amplitudes since
helicity is a good quantum number for Dirac spinors \cite{R04}.

Amongst the features that emerges for the reflected waves is the
 presence of both helicity {\it flip} and {\it non flip} terms. Furthermore,
the corresponding probabilities depend explicitly upon the
relative phase of the incoming helicity terms, when both are
present. This allows one to {\em  measure} this relative phase. As far as
we know this has not been noted previously.

Overall phase factors are of no physical significance and indeed cannot be measured in
any way. However, relative phases can be important and influence
experiments. We therefore start by recalling the significance of
some relative phases in diverse areas of particle physics \cite{R04,R0501,R0502}. The
relative phases that are most common in recent particle literature
are those between different but related decay amplitudes \cite{R0601,R0602,R0603,R0604}.
They are measurable when a multiple decay chain involves
alternative paths resulting in the same final states. They are
thus coherent and must be added before squaring. However, the
example in this paper involves the relative phases between
states, so let us limit our discussion to phases of this
type. Optical interference phenomena \cite{R0701,R0702,R0703,R0704} are the obvious
antecedents. A classic example is the two slit experiment with
possibly the inclusion of a transparent plate before one of the
slits to modify the relative phase between contributions \cite{R08}. Another
is that encountered in oscillation phenomena such as in
neutrino physics \cite{R0901,R0902}. In oscillations, two or more mass eigenstates
contained in a ``flavor" eigenstate \cite{R1001,R1002}, for kaons this flavor
would be the strong hypercharge \cite{R11}, develop differently in space and time
resulting in relative phases and consequent flavor superpositions. A third example stems
from an isospin calculation \cite{R12}. Let us determine, for example,
the ratio of cross sections
\begin{equation}
 \mbox{R}=\frac{\sigma(pp\to \pi^{\mbox{\tiny $+$}}\,d)}{\sigma(np\to\pi^{\0}\, d)}\,\,,
 \end{equation}
where $d$ is the singlet deuteron.
The incoming states, seen in an arbitrary Lorentz frame, formally
represent particle beams, even if a fixed target characterizes the
laboratory frame. There is no problem for the cross section in the
numerator, it involves pure $I=1$ states. But how do we represent
the ``np" beams in the denominator? The natural choice is the
 tensor product of the neutron and proton isospin kets, $|n\rangle\otimes |p\rangle$,
and this choice yields the result of $\mbox{R}=2$ \cite{R13} . However,
there is no \emph{operational distinction} between ``np" and ``pn",
although they are mathematical distinct, indeed they represent orthogonal states.
Now, if one uses ``pn",  $|p\rangle\otimes |n\rangle$,  instead of ``np", one
again obtains $\mbox{R}=2$.  Nevertheless, the most general beam-beam representation is
an admixture of both, including an arbitrary relative phase, i.e.
\begin{equation}
\label{state}
\cos\theta\,\,\,|n\rangle\otimes |p\rangle + e^{i\,\alpha}\, \sin\theta\,\,\,|p\rangle\otimes |n\rangle\,\,.
\end{equation}
Using
\[
|n\rangle\otimes |p\rangle  =  (\,|1\rangle + |0\rangle\,)\,/\,\sqrt{2}\,\,\,\,\,\,\,\mbox{and}\,\,\,\,\,\,\,
|p\rangle\otimes |n\rangle =  (\,|1\rangle - |0\rangle\,)\,/\,\sqrt{2}\,\,,\]
where $|0\rangle$ and $|1\rangle$ represent isospin $I=0$ and $I=1$ states, the superposition
 in Eq.(\ref{state}) can be written as follows
\begin{equation}
\left[\,(\cos\theta + e^{i\,\alpha}\, \sin\theta)\,\,\,|1\rangle +\,(\cos\theta - e^{i\,\alpha}\, \sin\theta)\,\,\,|0\rangle\,\right]\,/\,\sqrt{2}\,\,.
\end{equation}
In terms of the mixing angle $\theta$ and of the relative phase $\alpha$, the ratio of cross sections, R,
is then
\begin{equation}
\mbox{R} = 2\,/\,[\,1+ \,\cos\alpha\,\sin 2\theta\,]\,\,.
\end{equation}
This shows that $\mbox{R}=2$ is obtained not only for a pure $|n\rangle\otimes |p\rangle$
state or a pure  $|p\rangle\otimes |n\rangle$ state, but for a general linear combination of these
states if the relative phase is $\alpha=\pi/2$.

Finally, another type of relative phases are those between quark
  eigenstates used in the reduction of a $2\times 2$ unitary matrix to the Cabibbo
  rotation form \cite{R14}. A similar use is made in the reduction of the $3 \times 3$ unitary
  Kobayashi-Maskawa (KM) matrix to its standard form with three real
  angles and one complex phase \cite{R15}. The CP
  violating phase in the KM matrix is a very important phase but
  it is {\em not} a relative phase between two states, it appears
  between decay amplitudes \cite{R0601,R0602,R0603,R0604}.   The quark relative phases discussed here,
  although theoretically essential, are not experimentally measurable. This distinguishes
  them from the other examples given above.

This paper is structured as follows. In the next section, we introduce our formalism,
discuss the Dirac energy zones for  a stratified electrostatic potential, and give the general formulas of the plane wave functions in the free and potential regions.  In section III, by using a matrix approach, we explicitly calculate the reflection and transmission coefficients for planar scattering. In section IV, we study the
relative phase between the two helicity eigenstates of an
  incoming plane wave and analyze how it  modifies the probabilities of the reflected
  helicities. Our conclusions are drawn in the final section.

\section*{\normalsize II. FORMALISM}

We treat a potential $V_{\0}$ transforming as
 the zero component of a Lorentz four-vector and thus leading to
 the appearance of factors such as $E-V_{\0}$. The Dirac Hamiltonian in the
 presence of a constant electrostatic potential reads \cite{R0501}
 \begin{equation}
H_{_{D}}^{^{\mbox{\tiny (el)}}}=-\,i\,\boldsymbol{\alpha}\cdot \nabla + \beta\,m + V_{\0} \,\,.
 \end{equation}
  This can be contrasted for example with a
  scalar potential analysis
 in which the potential $V_s$ adds to the mass  leading to  factors such as
 $m+V_{s}$,
 \begin{equation}
H_{_{D}}^{^{\mbox{\tiny (sc)}}}=-\,i\,\boldsymbol{\alpha}\cdot \nabla + \beta\,\left(m + V_{s}\right) \,\,.
 \end{equation}
 There are multiple possible potential forms, including
Yukawa potentials \cite{R17}. None of these others will be discussed in
this work.

In a  previous study \cite{R0312}, for this potential, we showed that spin flip, which is
 rigorously absent in one dimensional scattering, is {\em present\,} in two
 dimensional or planar scattering  with the exception of ``head-on" collisions.
 Head-on collisions reduce the problem to the one dimensional case.
 We warn here of a possible confusion. For a
 head-on collision the reflected wave has an inverted momentum, so
 spin flip corresponds to helicity conservation. Thus absence of
 spin flip implies {\em total} helicity flip for head-on collisions.

 In the
 following, we use the Pauli-Dirac representation of the Dirac
 matrices. Since the scattering plane is chosen as the $x$-$y$ plane,
 the incoming particle four momentum reads
 \[ p_{_{\mbox{\tiny in}}}^{\,\mu} = \left(\,E\,,\,p_{\1}\,,\,p_{\2}\,,\,0\,\right)\,\,,\]
 with $E^{^{2}}=p_{\1}^{^{2}}+p_{\2}^{^{2}}+m^{\2}$. The reflected momentum is
 \[ p_{_{\mbox{\tiny ref}}}^{\,\mu} = \left(\,E\,,\,-\,p_{\1}\,,\,p_{\2}\,,\,0\,\right)\,\,.\]
 In the barrier region, $0<x<L$, the four momentum $q^{\mu}$  is obtained from the incident momentum by the substitution $E \to E-V_{\0}$ and observing that the discontinuity along the $x$-axis implies $q_{\2}=p_{\2}$,
 \[ q^{\mu} = \left(\,E-V_{\0}\,,\,q_{\1}\,,\,p_{\2}\,,\,0\,\right)\,\,.\]
The $q_{\1}$ is determined by the relation   $(E-V_{\0})^{^{2}}=q_{\1}^{^{2}}+p_{\2}^{^{2}}+m^{\2}$.
 In the potential region the solutions can be oscillatory in $x$ ($q_{\1}^{^{2}}>0$), this happens for the following energy zones
 \[
 \begin{array}{lclcl}
 &E &> \,\,V_{\0}+ \sqrt{p_{\2}^{^{2}}+m^{^{2}}} & & \mbox{(diffusion)}\,\,,\\
 m\,\,< & E& < \,\,V_{\0}- \sqrt{p_{\2}^{^{2}}+m^{^{2}}} & & \mbox{(Klein zone)}\,\,,
 \end{array}
 \]
  or evanescent ($q_{\1}^{^{2}}<0$), when
 \[
 V_{\0}- \sqrt{p_{\2}^{^{2}}+m^{^{2}}} \,\,< \,\,E \,\,< \,\,V_{\0}+ \sqrt{p_{\2}^{^{2}}+m^{^{2}}}\hspace*{1cm} \mbox{(tunneling)}\,\,.
 \]
The Klein energy zone for a step is consistent with Klein's
suggestion of the creation of particle-antiparticle pairs, with antiparticles propagating in the
potential region, which they see (because of their opposite
charge) as a potential well. In a previous work \cite{R0306}, we
have strongly argued that this phenomena (Klein pair production),
if it occurs for a step potential, must {\it necessarily} also
occur for a barrier. However, for simplicity we avoid this Klein
zone in this paper. As an aside, we note on the contrary, that
Klein pair production is {\em not} compatible with the kinematics
of a scalar potential.

Returning to our analysis, the outgoing (transmitted) momentum is
identical to the incoming (incident) momentum $p_{_{\mbox{\tiny out}}}^{\,\mu}=p_{_{\mbox{\tiny in}}}^{\,\mu}$.  The helicity operator is
 $\boldsymbol{\Sigma \cdot \mathcal{P}}/\,|\boldsymbol{p}\,|$, with  $\boldsymbol{\mathcal{P}}$ the three-momentum operator and $\boldsymbol{\Sigma}=\mbox{diag}[\,\boldsymbol{\sigma}\,,\, \boldsymbol{\sigma}\,]$.  The two planar eigenstates of this operator, with eigenvalues $\pm 1$,  are explicitly
 \begin{equation}
 \psi_{\pm}[\,p_{\1},E,x]\, \exp[\,i\,(p_{\2}\,y-E\,t)\,]\,\,,
 \end{equation}
 with
\[
\psi_{\pm}[\,p_{\1},E,x]=\frac{1}{2}\,\sqrt{\frac{E+m}{E}}\,\left(\,\,\pm\,
1\,\,,\,\,
\frac{p_{\1}+i\,p_{\2}}{\sqrt{E^{^{2}}-m^{\2}}}\,\,,\,\,
\sqrt{\frac{E-m}{E+m}}\,\,,\,\,
\pm\,\frac{p_{\1}+i\,p_{\2}}{E+m}\,\,
\right)^{^{\mbox{\footnotesize $t$}}}\,\exp[\,i\,p_{\1}\,x\,]\,\,.
\]
The plane wave solution is thus divided into three spatial
regions, with $\exp[\,i\,(p_{\2}\,y-E\,t)\,]$
 common to all three. For the  free potential region before the barrier,
region I with $x<0$,  we have
\begin{eqnarray}
\Psi_{\I}[\,p_{\1},E,x]&=&
I_{-}\,\psi_{-}[\,p_{\1},E,x]+I_{+}\,\psi_{+}[\,p_{\1},E,x]\, + \nonumber \\ & &
R_{-}\,\psi_{-}[\,-\,p_{\1},E,x]+R_{+}\,\psi_{+}[\,-\,p_{\1},E,x]\,\,,
\end{eqnarray}
where the $\pm\,1$ helicity amplitudes are indicated by
$I_{\pm}$ for the incoming  plane waves, moving from left to right
in the positive $x$ direction, and by $R_{\pm}$ for the reflected
waves, moving from right to left. For the barrier potential, region II
with $0<x<L$, we have
\begin{eqnarray}
\Psi_{\II}[\,q_{\1},E-V_{\0},x]&=&
A_{-}\,\psi_{-}[\,q_{\1},E-V_{\0},x]+A_{+}\,\psi_{+}[\,q_{\1},E-V_{\0},x] \,+ \nonumber \\
 & & B_{-}\,\psi_{-}[\,-\,q_{\1},E-V_{\0},x]+B_{+}\,\psi_{+}[\,-\,q_{\1},E-V_{\0},x]\,\,,
\end{eqnarray}
with $A_{\pm}$ and $B_{\pm}$ the helicity amplitudes for plane
waves travelling to and from the $x=L$ potential discontinuity.
Finally,  for the  free potential region after the barrier, region III with $x>L$, we have
\begin{eqnarray}
\Psi_{\III}[\,p_{\1},E,x]&=&
T_{-}\,\psi_{-}[\,p_{\1},E,x]+T_{+}\,\psi_{+}[\,p_{\1},E,x]\,\,,
\end{eqnarray}
with $T_{\pm}$ the helicity amplitudes for the outgoing plane
waves moving from left to right.

\section*{\normalsize III. REFLECTION AND TRANSMISSION COEFFICIENTS}

The continuity equations are
\begin{equation}
\Psi_{\I}[p_{\1},E,0]=\Psi_{\II}[q_{\1},E-V_{\0},0]\,\,\,\,\,\,\mbox{and}\,\,\,\,\,\,\,
\Psi_{\II}[q_{\1},E-V_{\0},L]=\Psi_{\III}[p_{\1},E,L]\,\,.
\end{equation}
The above continuity equations imply, after eliminating the $A$ and $B$ terms,
\begin{equation*}
\left[\,\begin{array}{c} I_{+}-I_{-}+R_{+}-R_{-}\\
\displaystyle{\frac{p_{\1}+i\,
p_{\2}}{\sqrt{E^{^{2}}-m^{\2}}}}\,(I_{+}+I_{-})+
\displaystyle{\frac{-p_{\1}+i\,
p_{\2}}{\sqrt{E^{^{2}}-m^{\2}}}}\,(R_{+}+R_{-})
\\ \displaystyle{\sqrt{\frac{E-m}{E+m}}}\, (I_{+}+I_{-}+R_{+}+R_{-})
\\
\displaystyle{\frac{p_{\1}+i\, p_{\2}}{E+m}}\,(I_{+}-I_{-})+
\displaystyle{\frac{-p_{\1}+i\, p_{\2}}{E+m}}\,(R_{+}-R_{-})
\end{array}
\right] = M\,\left[\,\begin{array}{c} T_{+}-T_{-}\\
\displaystyle{\frac{p_{\1}+i\,p_{\2}}{\sqrt{E^{^{2}}-m^{\2}}}}\,(T_{+}+T_{-})\\
\displaystyle{\sqrt{\frac{E-m}{E+m}}}\, (T_{+}+T_{-})
\\
\displaystyle{\frac{p_{\1}+i\,p_{\2}}{E+m}}\,(T_{+}-T_{-})
\end{array}
\right]\,\exp[i\,p_{\1}L]\,\,,
\end{equation*}
where
\[M=S\,D^{*}\,S^{^{-1}}\,\,,\]
with
\[S=\left(
\begin{array}{cccc}
1 & -1 & 1 & -1\\ \\
\displaystyle{\frac{q_{\1}+ip_{\2}}{\sqrt{(E-V_{\0})^{^{2}}-m^{\2}}}}
&
\displaystyle{\frac{q_{\1}+ip_{\2}}{\sqrt{(E-V_{\0})^{^{2}}-m^{\2}}}}
&
\displaystyle{\frac{-q_{\1}+ip_{\2}}{\sqrt{(E-V_{\0})^{^{2}}-m^{\2}}}}
&
\displaystyle{\frac{-q_{\1}+ip_{\2}}{\sqrt{(E-V_{\0})^{^{2}}-m^{\2}}}}\\
\\ \displaystyle{\sqrt{\frac{E-V_{\0}-m}{E-V_{\0}+m}}} &
\displaystyle{\sqrt{\frac{E-V_{\0}-m}{E-V_{\0}+m}}} &
\displaystyle{\sqrt{\frac{E-V_{\0}-m}{E-V_{\0}+m}}} &
\displaystyle{\sqrt{\frac{E-V_{\0}-m}{E-V_{\0}+m}}}\\ \\
\displaystyle{\frac{q_{\1}+ip_{\2}}{E-V_{\0}+m}} &
\displaystyle{-\,\frac{q_{\1}+ip_{\2}}{E-V_{\0}+m}} &
\displaystyle{\frac{-q_{\1}+ip_{\2}}{E-V_{\0}+m}} &
\displaystyle{-\,\frac{-q_{\1}+ip_{\2}}{E-V_{\0}+m}}
\end{array}
\right)
\]
and
\[D=\mbox{diag}[\,\exp(iq_{\1}L)\,,\,\,\exp(iq_{\1}L)\,,\,\,
\exp(-iq_{\1}L)\,,\,\,\exp(-iq_{\1}L)\,]\,\,. \]
After some algebra we find
\[ M = \left[\,\,
\begin{array}{llll}
a_{_{-}} & 0 & 0 & b_{_{+}}\\
0 & a_{_{+}} & b_{_{+}} & 0 \\
0 & b_{_{-}} &   a_{_{-}} & 0 \\
b_{_{-}} & 0 & 0 & a_{_{+}}
\end{array}
\right]\,\,,
\]
where
\[ a_{_{\pm}}=  \cos (q_{\1}L) \pm \displaystyle{\frac{p_{\2}}{q_{\1}}}\, \sin (q_{\1}L)\,\,\,\,\,\,\,\mbox{and}\,\,\,\,\,\, \, b_{_{\pm}} = -\,i\,\displaystyle{\frac{E-V_{\0}\pm m}{q_{\1}}}\, \sin (q_{\1}L)\,\,.
\]
Defining  $R$ and $\widetilde{R}$ as the reflected helicity
conserving and changing amplitudes and recalling that there is no
spin flip for the transmitted wave, which implies a unique
transmitted amplitude $T$, it can be proven that the incident,
reflected and transmitted amplitudes are
\begin{eqnarray*}
\mbox{\sc Incident} & : & \left(\,I_{+}\,,\,I_{-}\,\right)\,\,,\\
\mbox{\sc Reflected} & : & \left(\,I_{+}\,R+I_{-}\,\widetilde{R}\,,\,I_{-}\,R+I_{+}\,\widetilde{R} \,\right)=\left(\,R_{+}\,,\,R_{-} \,\right)\,\,,\\
\mbox{\sc Transmitted} & : & \left(\,I_{+}\,T\,,\,I_{-}\,T
\,\right)=\left(\,T_{+}\,,\,T_{-} \,\right)\,\,.
\end{eqnarray*}
The explicit expression for $R$, $\widetilde{R}$ and $T$ confirm
our previous published results and are explicitly
\begin{eqnarray}
\widetilde{R} & = & i\,\frac{m\,V_{\0}(p_{\1}+i\,p_{\2})}{q_{\1}
(E^{^{2}}-m^{\2})}\,
\sin (q_{\1}L)\,T\,\exp(i\,p_{\1}L)\,\,,\nonumber \\
R & = & -\,\frac{p_{\2}\,V_{\0}E\,(p_{\1}+i\,p_{\2})}{q_{\1} p_{\1} (E^{^{2}}-m^{\2})}\, \sin (q_{\1}L)\,T\,\exp(i\,p_{\1}L)\,\,,\\
T  & = & \exp(-\,i\,p_{\1}L)\,\,\mbox{\Large $/$}\,\left[\,\cos
(q_{\1}L) + i\, \frac{E\,V_{\0}-p_{\1}^{^{2}}}{p_{\1}q_{\1}}\,
\sin (q_{\1}L)\,\right]\,\,.\nonumber
\end{eqnarray}
They satisfy the necessary probability conserving feature
\begin{equation}
 |\,R\,|^{^{2}}+ |\,\widetilde{R}\,|^{^{2}} + |\,T\,|^{^{2}}=1\,\,.
\end{equation}
Again as in  previous articles we emphasize that our $R$, $\widetilde{R}$ and $T$ are amplitudes and not probabilities as sometimes defined in the literature. If the incoming plane wave is normalized conventionally then,
\[
|\,I_{+}\,|^{^{2}}+ |\,I_{-}\,|^{^{2}}=1
\]
and consequently probability conservation yields
\begin{equation} |\,R_{+}\,|^{^{2}}+ |\,R_{-}\,|^{^{2}} + |\,T_{+}\,|^{^{2}}+ |\,T_{-}\,|^{^{2}}=1\,\,.
\end{equation}
Obviously each reflected helicity amplitude $R_{\pm}$ is the sum of the conserved helicity  term $I_{\pm} R$ plus and helicity flipped contribution $I_{\mp} \widetilde{R}$. From the explicit expressions for $R$, $\widetilde{R}$ and $T$ we recognize the well known feature of resonance which occurs when
\[\sin(q_{\1}L)=0\,\,\,\,\,\Leftrightarrow\,\,\,\,\, q_{\1}L=n\,\pi\,\,\,(n=1,2,...)\,\,.\]
For these particular cases $R_{\pm}=0$ and $T_{\pm}=I_{\pm}$ (total transmission). It is worth recalling here that resonance phenomena is only valid for diffusion when coherence dominates. For finite spatial wave packets, coherence will break down if the barrier length $L$ grows sufficiently larger than the wave packet dimensions or for glancing impacts (high incident angles). In either case multiple reflected and transmitted wave packets appear at regular intervals. Resonance is then lost. It is also absent for tunneling phenomena but because of a different cause. Tunneling is consistent with the above formulas but with the significant difference of an {\it imaginary} value for $q_{\1}$. This follows from
\[ q_{\1}^{^{2}}=\left(\,E-V_{\0}\,\right)^{^{2}}-p_{\2}^{^{2}} - m^{\2}<0\]
since, for tunneling, evanescent $x$-waves exist in region II. In this case,
\[\sin (q_{\1}L) \to i\, \sinh(|q_{\1}L|)\]
and can never be zero.
 It is to be noted that since, by definition, for head on
collisions $p_{\2} =0$, the helicity conserving factor $R=0$.
Consequently, total helicity flip occurs for the reflection of head
on collisions.

\section*{\normalsize IV. RELATIVE PHASES ANALYSIS}

Consider, for the moment an incoming polarized beam say $I_{+}$, with $I_{-}=0$. This could be achieved via a Stern-Gerlach apparatus. Then from our results the transmitted wave will also be polarized
\[
T_{+}=I_{+}T\,\,\,\,\,\,\,\mbox{and}\,\,\,\,\,\,\,T_{-}=0\,\,,
\]
but this will not be the case for the reflected wave, indeed
\[
R_{+}=I_{+}R\,\,\,\,\,\,\,\mbox{and}\,\,\,\,\,\,\,R_{-}=I_{+}\widetilde{R}\,\,.
\]
However, these two helicity states will have a definite relative
phase since $\widetilde{R}/R$ is imaginary. The exact value of
this ratio depends upon momentum, but it does not depend upon the
values of $V_{\0}$ nor $L$, the barrier parameters. We thus have a
means of producing reflected waves with known relative phase and
calculable strengths.

Now, consider incoming waves, with a fixed relative phase. The
above reflected waves provides an example. Let
\[   I_{+}=|\,I_{+}\,|\, e^{i\,\alpha}\,\,\,\,\,\,\,\mbox{and}\,\,\,\,\,\,\,
I_{-}=|\,I_{-}\,|\, e^{i\,\beta}\,\,,\] so that the relative phase
of interest is $\alpha-\beta$. We subject this source to
scattering by our barrier. Again the transmitted waves are of
limited interest since they carry over the same relative phase
between $T_{+}$ and $T_{-}$. On the other hand the situation is
more complex for the reflected waves,
\begin{eqnarray}
|\,R_{+}\,|^{^{2}} & = & |\,I_{+}R\,|^{^{2}} + |\,I_{-}\widetilde{R}\,|^{^{2}} -\,
2\,|\,I_{+}I_{-}R\,\widetilde{R}\,|\, \sin (\alpha-\beta)\,\,,\nonumber \\
|\,R_{-}\,|^{^{2}} & = & |\,I_{-}R\,|^{^{2}} + |\,I_{+}\widetilde{R}\,|^{^{2}} +\,
2\,|\,I_{+}I_{-}R\,\widetilde{R}\,|\, \sin (\alpha-\beta)\,\,,
\end{eqnarray}
 which explicitly depend upon the relative phase. Thus, a second Stern-Gerlach apparatus
 followed by intensity (flux) measurement will yield the relative incoming phase.

 Our knowledge of relative phases is almost non existent. For example, in a broad particle
beam of constant density is the relative phase (when both
helicities are present) the same for all particles or is it
completely arbitrary? In the latter case the measurement of
relative phase as described above would yield a null mean value.
 Nevertheless, even in this latter case, the reflection of a \emph{polarized} beam will provide
 a fixed relative phase, for given incident angle and beam energy, for each individual particle
 in the beam. We believe these predictions are worthy of
 experimental verification.

\section*{\normalsize V. CONCLUSIONS}

In this work, we have studied the general planar diffusion (inclusive of
 tunnelling) of plane waves by an "electrostatic"
potential barrier. We have provided the explicit helicity
amplitudes for both  reflection and transmission. Since we consider
only plane waves we are in the limit of total coherence and this
is evidenced by the prediction of resonances in diffusion, when no
reflection occurs. We have confirmed some previous results, such
as the absence of spin flip for head-on collisions corresponding
to total helicity flip. There is a generalization of this result
($R=0$), for any incident angle (fixed $p_{2}/p_{1}$) the ratio
$R/\widetilde{R}$ tends to zero in the zero momentum limit. We
also confirm that the transmitted amplitudes are always
proportional to the incoming helicity amplitudes, no helicity flip
occurs.

The most interesting part of our results is that the probabilities
of the reflected helicity waves depend in a simple manner upon the
relative phase of the incoming helicity amplitudes. They, or at
least their average (if not constant),  may hence be determined
experimentally. Furthermore, if the incoming wave is polarized,
the reflected waves will have a fixed relative phase of $\pi/2$ and
modulus determined by the
scattering angle and the beam energy. Unfortunately, the apparatus described
in this paper does not furnish a method for creating helicity amplitudes with
any chosen relative phase (excluding $\pi/2$) but only for the determination
of any such phase.

As emphasized above these
results are valid only in the case of dominant coherence. However,
we have for diffusion a simple manner to confirm this situation by
identifying resonance phenomena. At certain incident angles or for
certain momenta or, for certain barrier lengths (with other
variables fixed) total transmission must occur or at least
dominate.

Thus any experimental test of our predictions must involve diffusion
of a particle beam of say electrons (initially polarized), by a stratified
electrostatic potential. After testing for coherence by means of resonance,
the first prediction to verify should be the absence of spin flip for head
on collisions. Subsequently, the explicit predictions for the transmitted
and reflected amplitudes can be tested. If our predictions are confirmed,
we suggest that a systematic study of the means of altering the incoming
relative phase, such as with the help of electromagnetic fields could be
attempted. It is very hard for us to suggest any practical
use of the above results. However, neither can we exclude any.
For example, we recall the significant multiple practical applications
of the related tunnel effect, such as in the electron microscope,
beyond its initial importance as an example of the failure of classical
physics.

We know very little  experimentally about relative phases. Here,
we have considered helicity amplitudes but they are of theoretical
interest also in iso-spin or other analysis. For example they may have
a significant effect  upon oscillation
formulae in particle physics. Finally, we recall that the formulas derived are
quite general and are valid both for diffusion (and Klein) and, when $q_{\1}$ is
imaginary, for tunneling. Integration over a convolution function of the incoming
momentum allows one to study wave packet sources but this generally alters the physics
involved, for example as we have already pointed out resonance phenomena no longer survive.

\section*{\small \rm ACKNOWLEDGEMENTS}

One of the authors (SdL) thanks the Department of Physics,
University of Salento (Lecce, Italy), for the hospitality and the
FAPESP (Brazil) for financial support by the Grant No. 10/02213-3.
The authors thank an anonymous referee for his comments and suggestions.

\end{document}